\documentclass[twocolumn,showpacs,prl]{revtex4}
\usepackage{graphicx}
\usepackage{dcolumn}
\usepackage{bm}

\begin{document}

\title{Superfluidity of fermions with repulsive on-site interaction in an anisotropic
optical lattice near a Feshbach resonance}
\author{B. Wang and L.-M. Duan}
\affiliation{FOCUS center and MCTP, Department of Physics,
University of Michigan, Ann Arbor, MI 48109}

\begin{abstract}
We present a numerical study on ground state properties of a
one-dimensional general Hubbard model (GHM) with particle assisted
tunnelling rates and repulsive on-site interaction (positive-U),
which describes fermionic atoms in an anisotropic optical lattice
near a wide Feshbach resonance. For our calculation, We utilize
the time evolving block decimation (TEBD) algorithm, which is an
extension of the density matrix renormalization group and provides
a well controlled method for one-dimensional systems. We show that
the positive-U GHM, when hole doped from half-filling, exhibits a
phase with coexistence of quasi-long-range superfluid and
charge-density-wave orders. This feature is different from the
property of the conventional Hubbard model with positive-U,
indicating the particle assisted tunnelling mechanism in GHM
brings in qualitatively new physics. \pacs{{03.75.Ss, 05.30.Fk,
34.50.-s}}
\end{abstract}

\maketitle

The combination of Feshbach resonance and optical lattice
techniques has opened up possibilities to investigate strongly
interacting ultracold atoms under tunable configurations \cite{1}.
Ability to control such strongly interacting systems provides an
unprecedented opportunity to explore interesting states of matter.
Many interesting physics has been predicted for ultracold atom
systems with fundamental Hubbard model hamiltonians. For example,
with Bose-Hubbard model and its derivations, people have studied
superfluid to Mott-insulator transition \cite{2}, existence of
supersolid order\cite{3}, etc, for ultracold bosons while with
Fermi-Hubbard model, Luther-Emery \cite{4} and FFLO \cite{5}
phases are predicted to be observable for ultracold Fermions with
attractive interaction. In particular, it is well known that for
repulsive (positive-$U$) conventional (Fermi-)Hubbard model the
susceptibility for superfluid and charge density wave (CDW) orders
are suppressed at low temperature and the leading quasi-long-range
order is given by a spin density wave (SDW) at any filling
fraction\cite{6}.

However, in this work, we show that coexistence of
quasi-long-range superfluid and CDW orders can be observed for
fermionic atoms with repulsive on-site interaction in an
anisotropic optical lattice near a wide Feshbach resonance. The
interactions in this strongly interacting system is described by a
one-dimensional positive-U general Hubbard model (GHM) with
particle assisted tunnelling rates \cite{8}. The GHM is an
effective one-band Hamiltonian that takes into account the
multi-band populations and the off-site atom-molecule couplings in
an optical lattice near a wide Feshbach resonance (see the
detailed derivation in Ref. \cite{8}). It is interesting to note
that the GHM with similar particle assisted tunnelling also arises
in different physical contexts, as proposed in Ref. \cite{9}. In
contrast with the case of conventional positive-U Hubbard model,
we show that the superfluid and CDW emerge as dominant quasi-long
range orders over spin orders for the positive-U GHM when the
system is significantly hole-doped below half-filling, although at
or very close to half-filling, the dominant correlation in GHM is
still anti-ferromagnetic. This feature indicates that the particle
assisted tunnelling in GHM brings in qualitatively new physics. It
makes the effective interaction in GHM doping dependent, showing
different behaviors with a possible phase transition in between.
We get our results through numerical calculation based on the time
evolving block decimation (TEBD) algorithm \cite{10,11}, which, as
an extension of the density matrix renormalization group method
\cite{12}, is a well controlled approach to deal with
one-dimensional systems. We compare our numerical results with
some known exact results for the conventional Hubbard model, and
the remarkably precise agreement shows that the calculation here
can make quantitatively reliable predictions.

As shown in Ref. \cite{8}, a generic Hamiltonian to describe
strongly interacting two-component fermions in an optical lattice
(or superlattice) is given by the following general Hubbard model:
\begin{eqnarray}
H &&= \sum_{i}\left[ Un_{i\uparrow }n_{i\downarrow }-\mu n_{i}\right]  \\
&&-\sum_{\left\langle i,j\right\rangle ,\sigma }\left[ t+\delta g\left( n_{i%
\overline{\sigma }}+n_{j\overline{\sigma }}\right) +\delta tn_{i\overline{%
\sigma }}n_{j\overline{\sigma }}\right] a_{i\sigma }^{\dagger
}a_{j\sigma }+H.c.  \nonumber
\end{eqnarray}%
where $n_{i\sigma }\equiv $ $a_{i\sigma }^{\dagger }a_{i\sigma }$, $%
n_{i}\equiv n_{i\uparrow }+n_{i\downarrow }$, $\mu $ is the
chemical potential, $\left\langle i,j\right\rangle $ denotes the
neighboring sites, and $a_{i\sigma }^{\dagger }$ is the creation
operator to generate a fermion on the site $i$ with the spin index
$\sigma $. The symbol $\overline{\sigma } $ stands for $\left(
\downarrow ,\uparrow \right) $ for $\sigma =\left( \uparrow
,\downarrow \right) $. The $\delta g$ and $\delta t$ terms in the
Hamiltonian represent particle assisted tunnelling, for which the
inter-site tunnelling rate depends on whether there is another
atom with opposite spin on these two sites. The particle assisted
tunnelling comes from the multi-band population and the off-site
atom-molecule coupling for this strongly interacting system
\cite{8}. For atoms near a wide Feshbach resonance with the
average filling number $\left\langle n_{i}\right\rangle \leq 2$,
each lattice site could have four different states, either empty
(with state $|0\rangle $), or a spin $\uparrow $ or $\downarrow $
atom ($a_{i\sigma }^{\dagger }|0\rangle $), or a dressed molecule
($d_{i}^{\dagger }|0\rangle $) which is composed by two atoms with
opposite spins. The two atoms in a dressed molecule can distribute
over a number of lattice bands due to the strong on-site
interaction, with the distribution coefficient fixed by solving
the single-site problem. One then can mathematically map the
dressed molecule state $d_{i}^{\dagger }|0\rangle $ to a double
occupation state $a_{i\downarrow }^{\dagger }a_{i\uparrow
}^{\dagger }|0\rangle $ by using the atomic operators $a_{i\sigma
}^{\dagger }$ \cite{8} . After this mapping, the effective
Hamiltonian is transformed to the form of Eq. (1). The GHM in Eq.
(1) reduces to the conventional Hubbard model when the particle
assisted tunnelling coefficients $\delta g$ and $\delta t$
approaching zero, as one moves far away from the Feshbach
resonance. Near the resonance, $\delta g$ and $\delta t$ can be
significant compared with the atomic tunnelling rate $t$ due to
the renormalization from the multi-band populations and the direct
neighboring coupling \cite{8}.

We consider in this work an anisotropic optical lattice for which
the potential barriers along the $x,y$ directions are tuned up to
completely suppress tunnelling along those directions. The system
becomes a set of independent one-dimensional chains. We thus solve
the GHM in one dimension through numerical analysis. For this
purpose, first we transfer all the fermion operators to the hard
core boson operators through the Jordan-Wigner transformation
\cite{13}. In the one-dimensional case, we can get rid of the
non-local sign factor, and after the transformation the hard core
boson operators satisfy the same Hamiltonian as Eq. (1). On each
site we then have two hard core boson modes which are equivalent
to a spin-$3/2$ system with the local Hilbert space dimension
$d=4$. We can therefore use the TEBD algorithm to solve this
pseudo-spin system \cite{10}. Similar to the density matrix
renormalization group (DMRG) method \cite{12}, the TEBD algorithm
is based on the assumption that in the one-dimensional case the
ground state $|\Psi \rangle =\sum_{i_{1}=1}^{d}\cdots
\sum_{i_{n}=1}^{d}c_{i_{1}\ldots i_{n}}|i_{1}\cdots i_{n}\rangle $
of the Hamiltonian with short-range interactions can be written
into the following matrix product form:
\begin{equation}
c_{i_{1}\ldots i_{n}}=\sum_{\alpha _{1},\ldots \alpha
_{n}=1}^{\chi }\Gamma _{\alpha _{n}\alpha _{1}}^{[1]i_{1}}\Gamma
_{\alpha _{1}\alpha _{2}}^{[2]i_{2}}\Gamma _{\alpha _{2}\alpha
_{3}}^{[3]i_{2}}\cdots \Gamma _{\alpha _{n-1}\alpha
_{n}}^{[n]i_{n}},
\end{equation}%
where $\Gamma ^{[s]i_s}$ denotes the matrix associated with
site-$s$ with the matrix dimension $\chi$. When $\chi =1$, the
assumption reduces to the mean-field approximation, and for a
larger $\chi$, the matrix product state well approximates the
ground state as it catches the right entanglement structure for 1D
systems \cite{10,11}. To use the TEBD algorithm, we just start
with an arbitrary matrix product state in the form of Eq. (2), and
evolve this state with the Hamiltonian (1) in imaginary time
through the propagator $e^{-Ht}$. The state converges to the
ground state of the Hamiltonian pretty quickly. From the final
ground state in the matrix product form, one can efficiently
calculate the reduced density operator and various correlation
functions. This calculation has a well controlled precision since
at each time step to update the matrix product state, the Hilbert
space truncation error can be suppressed by choosing an
appropriate matrix dimension $\chi $ \cite{10}. In this
calculation, we use the infinite lattice algorithm by assuming
that the lattice is bipartite and the ground state has a
translational symmetry for each sublattice \cite{10}. This allows
us to directly calculate the system in the thermodynamic limit.

To show that our calculation is capable of making reliable
predictions, we first test our results by comparing them with some
known exact results of the Hubbard model in certain cases. For the
Hubbard model at half-filling $\left\langle n_{i\uparrow
}\right\rangle =\left\langle n_{i\downarrow }\right\rangle =0.5$,
the ground state energy per site is known to have the analytic
expression $E=-4\int_{0}^{\infty }\frac{J_{0}(\omega )J_{1}(\omega
)d\omega }{\omega \lbrack 1+\exp (\omega U/2)]}$ in the
thermodynamic limit from the exact Bethe ansatz solution
\cite{14}, where $J_{0}$ and $J_{1}$ are Bessel functions and we
have chosen the tunnelling rate $t$ as the energy unit. In Fig. 1
(a), we show our numerical results for the ground state energy of
the Hamiltonian (1) with $\delta g=\delta t=0$, and one can see
that it agrees very well with the exact energy of the Hubbard
model in particular when $U > t$. The error is in general smaller
than $10^{-3}$ as shown in Fig. 1(b). In this and the following
calculations, we choose the matrix dimension $\chi =40$. We have
tried larger $\chi $ which gives better precision, but we choose
$\chi =40$ to have a faster speed and its precision is enough for
our purpose.

\begin{figure}[tbp]
\includegraphics [height=4.5 cm, width=8 cm]{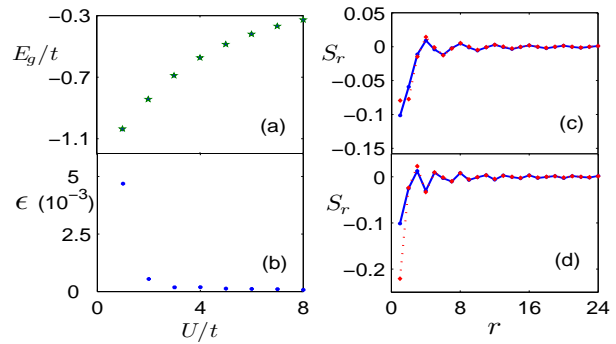}
\caption[Fig.1]{The numerical result for the Hubbard model
compared with some known exact result. (a) Ground state energy as
a function of $U$ at half-filling (energy in the unit of $t$),
where data points marked by solid dots are from the exact Bethe
ansatz solution while those marked by pentagram are from our
numerical program; (b) The relative error in the ground state
energy; (c) Real-space spin correlation function at the filling
fraction $\langle n_i\rangle=0.5$ and $U=8t$, compared with the
asymptotic form in Eq. (3) (solid curve) with $K_{\rho}=0.62$ and
$A=0.13$. (d) Similar to (c), except that $\langle
n_i\rangle=0.75$, and the corresponding $K_\rho=0.60$, $A=$0.19.}
\end{figure}

We have also tested the final state from our calculation by
comparing its correlation functions with some known results. It is
difficult to get correlations analytically from the Bethe ansatz
solution, but from the bosonization approach to the
one-dimensional Hubbard model, we know its correlation functions
take certain asymptotic forms. For instance, one can look at the
spin-spin correlation, defined as $S_{r}\equiv \langle
\mathbf{s}_{i}\cdot \mathbf{s}_{i+r}\rangle$, where the spin
operator for the site $i$ is given by $\mathbf{s}_{i}\equiv
a_{i\alpha }^{\dagger }\mathbf{\sigma }_{\alpha \beta }a_{i\beta
}/2$ with $\alpha $ and $\beta =\downarrow $, $\uparrow $ and
$\mathbf{\sigma }$ standing for the Pauli matrices. The
correlation $S_{r}$ is independent of $i$ because of the
translational symmetry. The Hubbard spin correlation function has
the following asymptotic form \cite{15}
\begin{equation}
S_{r}=-\frac{1}{(\pi r)^{2}}+\frac{A}{r^{1+K_{\rho }}}\cos
(2k_{F}r)\ln ^{1/2}(r)+...,
\end{equation}
where $K_{\rho }$ is the Luttinger parameter whose value has been
determined before from the exact Bethe ansatz solution
\cite{14,15}, $k_{F}$ is the fermi momentum related to
the filling number $\left\langle n_{i}\right\rangle $ through $%
k_{F}=\left\langle n_{i}\right\rangle \pi/2$, and $A$ is a
non-universal model dependent constant. In Fig.1 (c) and (d), we
compare our calculation results for $S_{r}$ with this asymptotic
form for filling number $\langle n_{i}\rangle =0.5$ and $0.75$,
and the agreement is again remarkable as long as $r$ is not too
small (the expression of $S_{r}$ in Eq. (3) is not accurate for
small $r$).

\begin{widetext}

\begin{figure} [tbp]
\includegraphics [height=4 in, width=6.5 in]{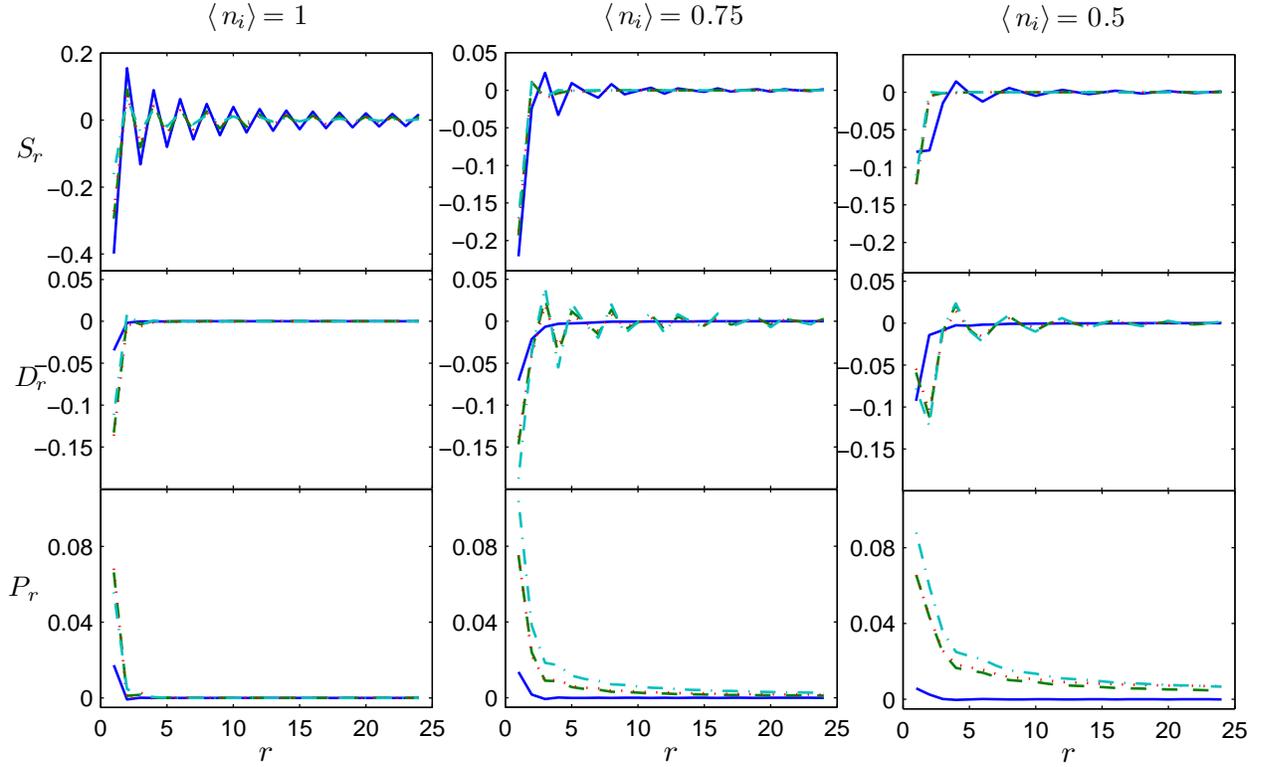}
\caption[Fig.2] {The numerical results for the spin ($S_r$), the
CDW ($D_r$), and the pair ($P_r$) correlation functions for the
GHM with different particle assisted tunnelling rates and at
different filling fractions, where $\delta g=0$ $\delta t=0$ for
solid curves, $\delta g=3t$ $\delta t =-6t$ for dashed curves,
$\delta g=3t$ $\delta t=-3t$ for dotted curves, and $\delta g=7t$
$\delta t=-14t$ for dash-dotted curves.}
\end{figure}

\end{widetext}

With the confidence in numerics built from the above comparison,
we now
present our main calculation results for the repulsive GHM in Eq. (1) with $%
U>0$. Apart from the spin correlation $S_{r}$ defined before, we
also calculate the charge-density-wave (CDW) correlation, defined
as $D_{r}\equiv \langle n_{i}n_{i+r}\rangle -\langle n_{i}\rangle
\langle n_{i+r}\rangle $, and the pair (superfluid) correlation,
defined as $P_{r}\equiv \langle a_{i\uparrow }a_{i\downarrow
}a_{i+r\downarrow }^{\dagger }a_{i+r\uparrow }^{\dagger
}\rangle $. The results are shown in Fig. 2 for different filling fraction $%
\langle n_{i}\rangle $ and for models with different particle
assisted
tunnelling rates $\delta g$ and $\delta t$. First at half filling with $%
\langle n_{i}\rangle =1$, the correlation functions $S_{r}$, $D_{r}$, and $%
P_{r}$ for the GHM with different $\delta g$ and $\delta t$ all
look qualitatively similar to the corresponding results for the
conventional Hubbard model, although with increase of the
coefficient $\delta g$ the spin correlation reduces a bit while
the CDW and superfluid correlations increase slightly. Clearly,
the dominant correlation in this case is in spin which suggests a
quasi-long range anti-ferromagnetic order. In this and following
calculations, we take $U=8t$ for all the cases, which corresponds
to a significant on-site repulsion.

Qualitatively different results show up when the system is doped
with holes. At the filling fraction $\langle n_{i}\rangle =0.75$,
although for the Hubbard model the spin correlation is still the
dominant one (the spin density wave order has been pinned to the
corresponding $2k_{F}=3\pi /4$), for the GHM\ with a noticeable
$\delta g$, the superfluid and the CDW emerge as the leading
quasi-long-range orders, and their correlations increase
significantly and decay much slower in space compared with the \
spin correlation when $\delta g$ grows. These features become more
evident when we further increase the doping. For instance, at the
right column of Fig. (2), we show the correlations for the filling
fraction $\langle
n_{i}\rangle =0.5$. The qualitative behavior is similar to the case with $%
\langle n_{i}\rangle =0.75$, but the CDW and superfluid
correlations for the GHM\ get significantly larger at long
distance, and the contrast with the Hubbard model becomes sharper.
One also note that for all these calculations, change of the
coefficient $\delta t$ in the GHM\ makes little difference to the
result. This is understandable as a significant positive $U$
suppresses the possibility of double occupation in the lattice,
and the $\delta t$ term in the GHM has no effect without double
occupation. The $\delta g$ term in the GHM, however, is critically
important, which favors superfluidity in general and brings in the
qualitatively different features mentioned above.

\begin{figure}[tbp]
\includegraphics [height=6 cm, width=8 cm]{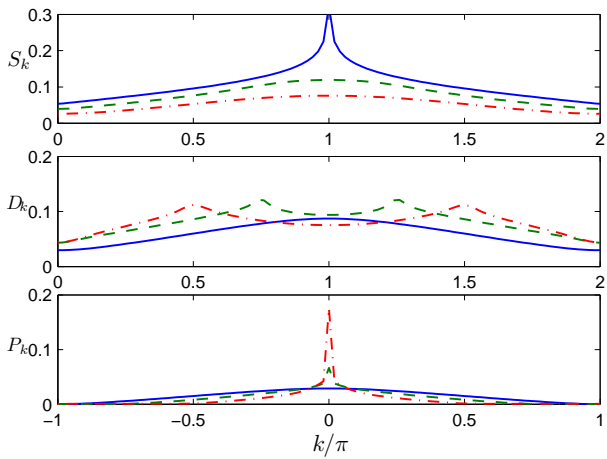}
\caption[Fig.3]{The spin, the CDW, and the pair correlation
functions in momentum space for the GHM with $\delta g=3t$ $\delta
t=-3t$. The solid, dashed, and dash-dotted curves correspond to
filling factor $\langle n_{i}\rangle =1$, $0.75$, and $0.5$,
respectively. For the calculation of the Fourier transformation,
we have used the real space correlation functions for $N=100$
sites.}
\end{figure}

To show the spatial structure of these quasi-long range (QLR)
orders, we plot in Fig. 3 the spin, the CDW, and superfluid
correlations in the momentum space for the GHM with $\delta g=3t$
at different filling fractions $\langle n_{i}\rangle $. The
momentum space correlations are defined by the Fourier transform
$X_{k}=1/\sqrt{N}\sum_{r=0}^{N}X_{r}\cos (kr)$, where $X$ stands
for the correlations $S$, $D$, or $P$. From these momentum space
curves, one can clearly see that this GHM at half filling has a
QLR anti-ferromagnetic order (characterized by the peak at $k=\pi
$), and away from half filling a QLR superfluid order (peak at
$k=0$) and CDW order (peaks at $k=2k_{F}$ and $2\pi -2k_{F}$,
where $2k_{F}=3\pi /4$ ($\pi /2)\ $for the filling fractions
$\langle n_{i}\rangle =0.75$ ($0.5$), respectively). The peaks in
Fig. 3 have finite widths because these orders in 1D are only
quasi-long-range with algebraic decay. Note that if we turn on
small tunnelling interaction between different 1D chains, a
leading QLR order, such as superfluid order, could be stabilized
to a true long range order \cite{6}. The GHM thus provides an
example of a microscopic Hamiltonian that with hole doping from
half filling, an anti-ferromagnetic phase could be transferred to
a superfluid phase (or a CDW phase in some case depending on which
order becomes more dominant with the inter-chain coupling). The
correlations that characterize these QLR orders can be detected
for the cold atomic gas, for instance, through the method
described in Ref. \cite{17}.

In summary, we have investigated the ground state properties of
the general Hubbard model with repulsive on-site interaction in
one dimension through well controlled numerical analysis. For the
system with significant particle assisted tunneling rates $\delta
g$ and $\delta t$, we have found coexistence of quasi-long range
superfluid and charge-density-wave orders when the system is
hole-doped from half filling. This feature is in sharp contrast
with convention Hubbard model, in which case for positive-$U$ the
charge and superfluid orders are always suppressed regardless of
the filling fraction. With a combination of the Bosonlization
approach and the numerical method here, it may be possible to
determine the compete phase diagram for the GHM. The model here
describes strongly interacting fermionic atoms in an anisotropic
optical lattice. The possibility of a transition from an
anti-ferromagnetic phase to a superfluid phase for the GHM with
hole doping may also have interesting indications for other areas.

This work is supported under the MURI program, and under ARO Award
W911NF0710576 with funds from the DARPA OLE Program.

\end{document}